# A UML Profile for Bitcoin Blockchain

**Behrouz Sefid-Dashti[1,*], Javad Salimi Sartakhti[2,*], Hassan Daghigh[3]**

[1]*Department of Computer Engineering, University of Kashan, Kashan, Iran,* b.sefiddashti@grad.kashanu.ac.ir
[2]*Department of Computer Engineering, University of Kashan, Kashan, Iran,* salimi@kashanu.ac.ir
[3]*Faculty of Mathematical Science, University of Kashan, Kashan, Iran,* hassan@kashanu.ac.ir
[*]*Corresponding Author*

## Abstract

Blockchain has received attention for its potential use in business. Bitcoin is powered by blockchain, and interest in it has surged in the past few years. It has many uses that need to be modeled. Modeling is used in many walks of life to share ideas, reduce complexity, achieve close alignment of one person's viewpoint with another and provide abstractions of a system at some level of precision and detail. Software modeling is used in Model Driven Engineering (MDE), and Domain Specific Languages (DSLs) ease model development and provide intuitive syntax for domain experts. The present study has designed and evaluated a meta-model for the bitcoin application domain to facilitate application development and help in truly understanding bitcoin. The proposed meta-model, including stereotypes, tagged values, enumerations and a set of constraints defined by Object Constraint Language (OCL), was defined as a Unified Modeling Language (UML) profile and was implemented in the Sparx Enterprise Architect (Sparx EA) modeling tool. A case study developed by our meta-model is also presented.

**Keywords**, Meta-model, UML profile, Bitcoin, Blockchain, OCL, Domain-specific language

## 1. Introduction

Blockchain provides enormous advantages. In light of the new computing paradigm, there is a need for developing blockchain-specific software engineering best practices [1] and modeling profiles. Modeling is a prerequisite for relevant methodologies and reference architectures [2]. Models are used in model-driven engineering (MDE), provide feedback on model's correctness prior to development [3], help reduce complexity, focus on software development goals, leave other aspects aside. In addition, Domain-specific languages (DSLs) provide domain-specific primitives which not only ease model development but also present intuitive syntax for domain experts, and the possibility of code generation for narrow domains [4]. Moreover, complementary models might be used to provide important insights into some complex phenomena. *Unified Modeling Language (UML)* is a graphical language standardized by the *Object Management Group (OMG)* for modeling software-intensive systems [5], and is executable, at least in part

[6]. While different UML diagrams are complementary and appropriate for different aspects of software systems, UML provides extensibility mechanisms to define meta-models known as UML profiles [7-8]. A profile is a lightweight mechanism to extend the UML standard [7] and defines domain-specific or platform-specific elements, connectors, and diagrams which aim to facilitate the modeling of applications for people interested in that domain or platform. Various studies have been carried out on blockchain modeling. For example, Rocha and Ducasse [9] used a UML class diagram to model smart contracts; they used a class with "chain" icon to represent the smart contracts, but they used neither stereotyping (i.e. stereotypes, tagged values, and constraints) nor comprehensive combinations of connectors (e.g., dependencies, composition, and aggregation). Bollen [10] applied fact-based modeling to provide conceptual model of Hyperledger Fabric using *fact definition* and *rule validation,* but his model requires a significant cognitive load to cross through annotated diagrams and the provided tables. The integration and orchestration [11] of blockchain-specific services with other organizational services have also been studied [12]. Vingerhouts et al. [12] used i* modeling notation and UML Use Case and Sequence diagrams for requirement engineering and modeled contract interactions in lieu of detailed design. Despite these attempts at blockchain modeling, most of them are modeling instead of meta-modeling and to the best of our knowledge, there is no bitcoin-specific meta-model supported by tools, and this study is the first one to have aimed at assisting application development using profiling.

This study aims to build on extant research to design and evaluate a new viable meta-model. Hence, in this study, we build on UML, OCL, and bitcoin's ontology to propose a UML profile which aligns models with bitcoin's ontology and automate evaluation of conformance. The former is achieved by defining relevant stereotypes, tagged values, and enumeration inside our profile, while the latter is achieved by defining relevant OCL constraints. Our profile has been implemented in the Sparx EA[1] modeling tool.

The remainder of this paper has been structured as follows: Section 2 briefly introduces some concepts upon which our approach is built. Our proposed UML profile is presented in Section 3. In the next section, a case study investigated and modeled by the proposed UML profile is presented. Finally, the paper is summed up in Section 5, and conclusions are provided.

## 2. Background

This section briefly reviews the most important concepts used for the development of our UML profile. The first sub-section briefly overviews bitcoin blockchain, and the second sub-section introduces UML extensibility mechanisms.

---

[1] www.sparxsystems.com

## 2.1. Bitcoin Blockchain

Blockchain is a distributed ledger without a central node of control. It keeps cryptographically signed, irrevocable, recorded events (i.e. transactions) and is shared across participants which can independently store, verify, and audit information over a peer-to-peer network. Each transaction is referenced to previous transactions, and each block keeps some transactions and uses a hash value to reference to a previous block, so that blockchain is auditable, and each transaction can be traced. When the majority of participants verify and add a block of transactions to their local instance of blockchain, the content of that block will be immutable [13, 14].

Different types of blockchain exist. A public blockchain allows anyone to join, and participants are anonymous with equal rights to access and validate transactions. A private blockchain is controlled by a single organization which decides on membership and determines the roles that each node can play in the blockchain. A consortium blockchain, aka federated blockchain, is controlled by a group of organizations and helps them find solutions together. A hybrid blockchain is a combination of a private and a public blockchain. It is controlled by a single organization, and transactions are made and stored privately, but they can be made public and be verified by public blockchain members. Finally, emerging fourth and fifth generation blockchains use microprocessors, mobile devices, and Artificial Intelligence (AI) to improve security and scalability, and assist mining by features such as AI based consensus algorithms. Bitcoin is powered by a public blockchain. The rest of this sub-section elaborates the bitcoin blockchain.

Bitcoin is backed by *mining* through significant computational efforts and *consensus* through the commitment of the majority of blockchain nodes. *Mining* is a process that increases the security of bitcoin by introducing the computational effort required for adding a new block to the blockchain [15]. Miners are rewarded with new coins generated in each new block, and they receive a transaction fee from each transaction included in that new block.

Achieving these rewards requires computing a cryptographic hash function several times, until a miner finds a block hash value less than a predefined threshold entitled *Target Value*. The resultant solution is called *hashcash*, which is a *Proof-of-Work (PoW)* system invented by Adam Back in 1997 [16] that ensures that issued coins are backed by significant computational efforts. The computed hash and all contained transactions are evaluated by all nodes of the bitcoin network, and each node adds the confirmed block to its local instance of the blockchain. When the majority of nodes participating in the bitcoin network add the new block to their local instance of blockchain, consensus is achieved.

Miners use *Nonce* to compute hash values. *Nonce* is a random 32-bit (4-byte) number stored in a bitcoin block header to achieve a block hash that is smaller than *Target Value* (i.e. a hash value which contains enough leading zeros). *Target Value* is used to adjust mining difficulty. In addition to the 4-byte nonce, miners can use 8 bytes of the *coinbase transaction* as an *extra nonce* field [15].

There are different types of mining. A miner may do his task of mining operations with no help (i.e. solo mining), use cloud resources (i.e. cloud mining), or join a mining pool constructed form a large number of miners (i.e. pooled mining). These types of mining use different protocols which are included in our proposed meta-model.

Bitcoin blockchain stores financial transactions and defines a coin as a chain of digital signatures [17]. Each transaction encodes the transfer of money between participants and includes at least one input, at least one output, and a transaction fee. An exception in this case is the first transaction of each block, which is a special transaction that generates a new coin and is called *coinbase transaction* [15].

A *transaction output* determines the number of Satoshis to be transferred and provides a *locking-script* which indicates the next owner of the coin(s). A transaction output can be locked by any equation defined by the *Script* language of bitcoin [15], but most of them (i.e. *Pay-to-Public-Key (P2PK)*, *Pay-to-Public-Key-Hash (P2PKH)*, *Multi-Signature (multisig)*, and *Pay-to-Script-Hash (P2SH)* [15]) use an *Elliptic Curve Digital Signature Algorithm (ECDSA)* as proof of ownership [18].

A *transaction input* consists of two fields: the address of an *Unspent Transaction Output (UTXO)* and an *unlocking-script*, which is checked against the *locking-script* of the referenced transaction output.

A *transaction fee* is collected by mining nodes. The fee is not explicitly stated in a transaction and is defined as the difference between the sum of transaction inputs and the sum of transaction outputs [15].

Finally, bitcoin wallets manage the wallet owners' key sets (including private keys, public keys, child keys, seeds, and mnemonic code words [15]), provide transparency regarding the blockchain access, and are the source for creating new transactions. They scan the blockchain to detect the UTXOs which are spendable by the wallet owner's private keys. Whenever a user needs to spend some bitcoins, their wallet selects a combination of their scanned UTXOs, provides an appropriate unlocking script for each of these selected UTXOs, and determines the transaction fee.

## 2.2. UML Profiling

Different notations and meta-modeling frameworks [19, 20, 21] and extension of these meta-models [4, 8, 22, 23, 24] exist. Meta Object Facility (MOF) [20] is a well-accepted framework developed by OMG, and approaches model development in multiple levels of abstraction (e.g., a four-layered metamodel architecture including meta-meta-models, meta-models, models, and user objects). Subsequent layers allow navigation from an instance to its meta-object (its classifier) and vice versa. UML itself has been defined by MOF. Profiles are lightweight mechanisms to extend the UML standard [7]. This section describes UML profiling and related concepts which are used to define our proposed profile for bitcoin.

A profile is a mechanism for customizing UML to meet the needs of a particular context. A profile can be defined for a platform that is being targeted (e.g., Java EE or .NET) or a domain with which one is working (e.g., financial or blockchain) [25]. UML profiles define a concise dialect of UML for a specific family of applications (e.g., UML profile for wireless sensors [8], electronic and electrical waste [22], big data [26], aerospace systems safety [27], hazard mitigation [23], and publish/subscribe paradigm [24]) and are composed of *Stereotypes*, *tags*, and *constraints* [28].

Stereotypes explicitly specify that an element has a special intent or role in a model [25, 28]. A stereotype is shown using guillemots at either end of the stereotype name, as in «stereotype_name». However, they can be substituted by angle brackets, as in <<stereotype_name>>. In Sparx EA, stereotype definition and implementation are denoted by specifying the name of the stereotype between and before two angle brackets, as in <<stereotype_name>> and stereotype_name <<>>, respectively.

Tagged values define information needed by a stereotype to perform its responsibilities [28]. They are meta-attributes which show some properties of model elements such as stereotypes [7]. To encourage reuse, UML 2.0 restricted declaration of *stereotypes* and *tagged values* to UML profiles [25].

Finally, constraints restrict model elements. They are defined in the profile but evaluated in the model. Constraints are defined by OCL, a formal yet easy-to-understand expression language for specifying constraints, which allows values to be checked but not changed [25]. OCL is helpful in creating the metamodel of a language, which is a description of all the concepts that can be used in that language and includes all meta-classes of that language and the relationships between them [29].

Figs. 1 and 2 show an example of stereotype definition and usage, respectively. In the first figure (Fig. 1), an `extension` arrow with a solid arrowhead pointing from `ExampleStereotype` stereotype to `Class` meta-class depicts that the `ExampleStereotype` stereotype which is tagged with `ExampleTaggedValue` tagged value can be applied to classes. Fig. 2 depicts a stereotyped class containing an attribute, an operation, and two tagged values. This example also includes an OCL constraint. The mentioned OCL invariant states that the `firstAttribute` attribute of `ExampleClass` class must be greater than zero.

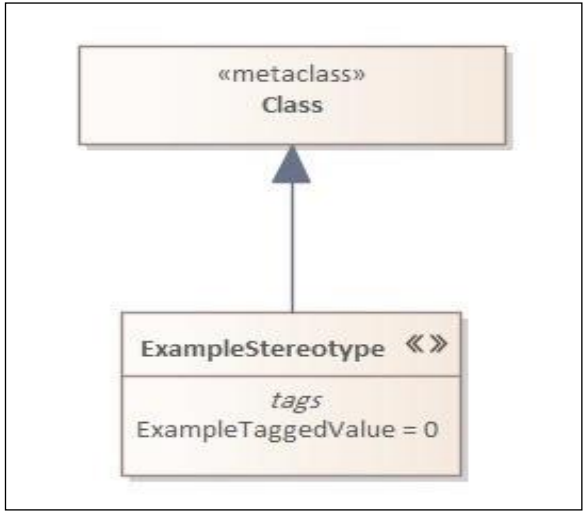 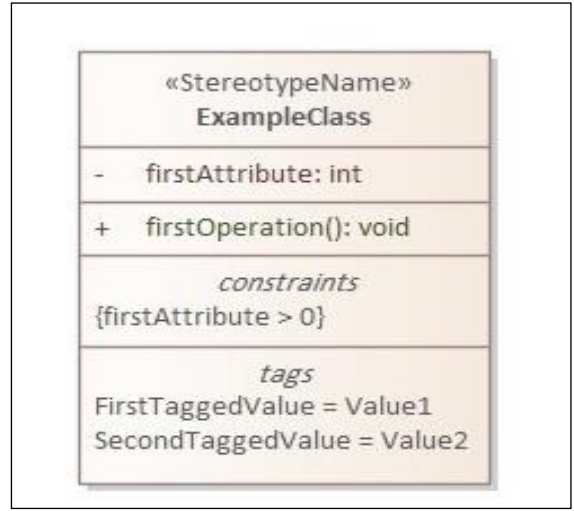

**Fig. 1.** An example of stereotype definition  **Fig. 2.** An example of stereotype usage

In this paper, the *stereotypes*, *tags*, and *constraints* are represented by UML standard notation, and the proposed UML profile was implemented using *Model Driven Generation (MDG) technologies* [30] feature of Sparx EA modeling tool. We defined the appropriate OCL constraints on our proposed metamodel to allow greater integrity of application models.

## 3. Proposed Meta-Model

This section describes our UML profile and its implementation in Sparx EA modeling tool. The proposed profile is called Bitcoin Modeling (BitML) and is defined using UML notation. This notation is common for profile definition with some exceptions that are considered irrelevant.

### 3.1. Profiling Definition

BitML includes two types of diagrams, i.e. a *Transaction Processing* diagram and a *Network* diagram. As profiles extend UML, our proposed diagrams are intended to be used along with UML standard diagrams. For example, BitML stereotypes can be used in both a *Transaction Processing* diagram and a UML sequence diagram to show both application structure and its runtime behavior.

A *Transaction Processing* diagram is defined as an extension of UML class diagram and provides stereotypes, tagged values, enumerations, and constraints to model on-chain and off-chain transactions in a way consistent with bitcoin ontology. *Network* diagram is defined as an extension of the UML deployment diagram and provides stereotypes, tagged values, enumerations, and constraints to model different bitcoin node types and protocols. Figs. 3 and 4 depict profile elements (i.e. stereotypes, tagged values, and enumerations) and an exemplar set of constraints on the proposed UML profile, respectively. Furthermore, Figs. 5 and 6 depict how the proposed connectors (i.e. *Spend* connector, *Uplock* connector,

*PBKDF2KeyStretching* connector, *HMAC-SHA512* connector, and *RIPEMD160HashOfSHA256Hash* connector) are defined and constrained, respectively.

As shown in Figs. 3 and 5, the profile is composed of 42 stereotypes (16 for classes, 14 for attributes, eight for operations, and five for connectors), 23 tagged values, six datatypes, and a set of constraints. Proposed meta-model includes the following enumerations:

I. *ScriptType*: This enumeration defines five values.
   1. Pay-to-Public-Key (P2PK)
   2. Pay-to-Public-Key-Hash (P2PKH)
   3. Pay-to-Script-Hash (P2SH)
   4. Pay-to-Witness-Public-Key-Hash (P2WPKH)
   5. CustomScript.

   The first four values correspond to the most common script types in bitcoin. The last value shows the custom script development [15] capability which made bitcoin extendable and a programmable money.

   *TransactionPosition*: A transaction may be on-chain or off-chain which will be executed on or out of the blockchain respectively. An off-chain transaction will be executed out of the blockchain, but its execution is bonded to some blockchain data, and its execution results will be saved on the blockchain too. This enumeration includes two values for on-chain and off-chain transactions.

   *PayToScriptHashType*: Regarding Pay-to-Script functionality, a mining node may support either Pay-to-Script-Hash (P2SH) or CheckHashVerify (CHV). P2SH or CHV correspond to BIP-16 or BIP-17, respectively.

   *CommunicationProtocol*: Bitcoin nodes communicate over the Bitcoin P2P protocol. In addition, some miners and mobile wallets communicate over the Stratum protocol. Finally, a pool miner may communicate over a specialized mining pool protocol. This enumeration defines these three values.

   *HashCashFucnctionType* **and** *HashCashVerificationFunctionType*: These enumerations correspond to Hashcash. As mentioned, Hashcash is a PoW system. Depending on the hash function that is used, there are three Hashcash variants (i.e. SHA-1, scrypt hash function, and double SHA-256). Bitcoin uses Hashcash with double SHA256 hash.

Depicted in Fig. 3, our profile includes 42 stereotypes, sixteen of which (*BitcoinNode, Block, BlockHeader, Transaction, TransactionOutput, AbstractTransactionInput, TransactionInput, CoinbaseTransactionInput, LockingScript, UnlockingScript, EllipticCurveSignature, MnemonicCodeWord, Seed, PrivateKey, PublicKey,* and *PublicAddress*) extend UML class. Two types of transaction inputs are abstracted by the *AbstractTransactionInput* abstract class. The *UML Composition* relationship between *AbstractTransactionInput* and *Transaction* stereotypes indicates that each transaction requires at least one transaction input. Depicted by *UML Generalization*, *TransactionInput* and *CoinbaseTransactionInput* stereotypes are specializations of *AbstractTransactionInput* stereotype. The former represents a normal transaction input which includes three attributes that point to an UTXO which will be unlocked by an instance of a class stereotyped as *UnlockingScript*. The latter represents a coinbase input. Coinbase is the input of coinbase transaction which is the first transaction of each block and generates new coins as mining rewards. The *MinerPay2ScriptHashStandard* tagged value of the *CoinbaseTransactionInput* stereotype uses *PayToScriptHashType* enumeration to specify either P2SH or CHV, which is supported by the miner. Coinbase transaction input stores a block height and possibly an – extra-nonce value which correspond to *BlockHeight* and *ExtraNonce* stereotyped attributes. Furthermore, the *Spend* connector, which is defined in Fig. 5 and constrained in Fig. 6, applies to instances of classes stereotyped as *TransactionOutput*, *AbstractTransactionInput*, *TransactionInput*, and *CoinbaseTransactionInput,* showing that a transaction output may spend either a coinbase transaction or a normal transaction.

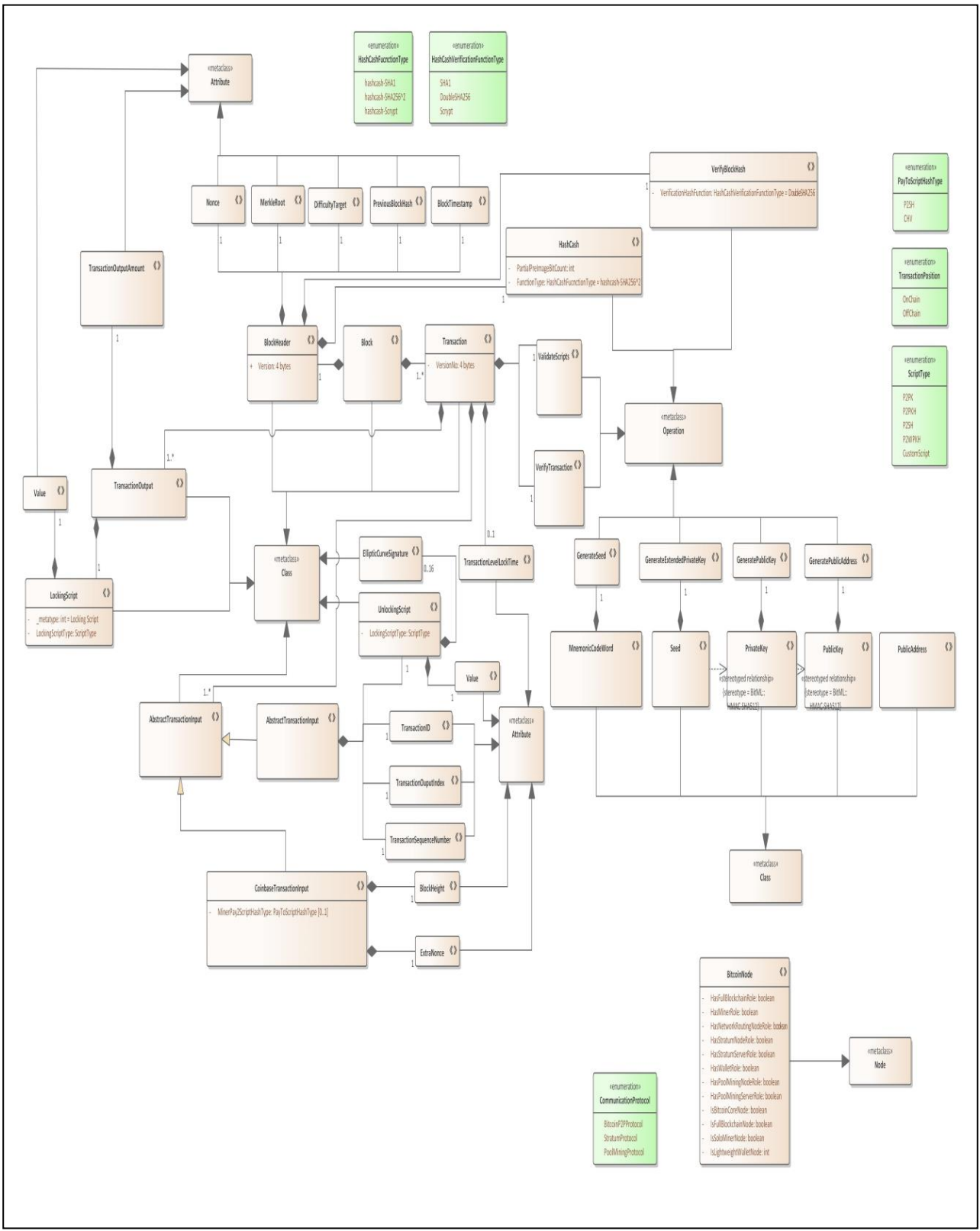

**Fig. 3.** Proposed UML profile, except for connectors meta-model

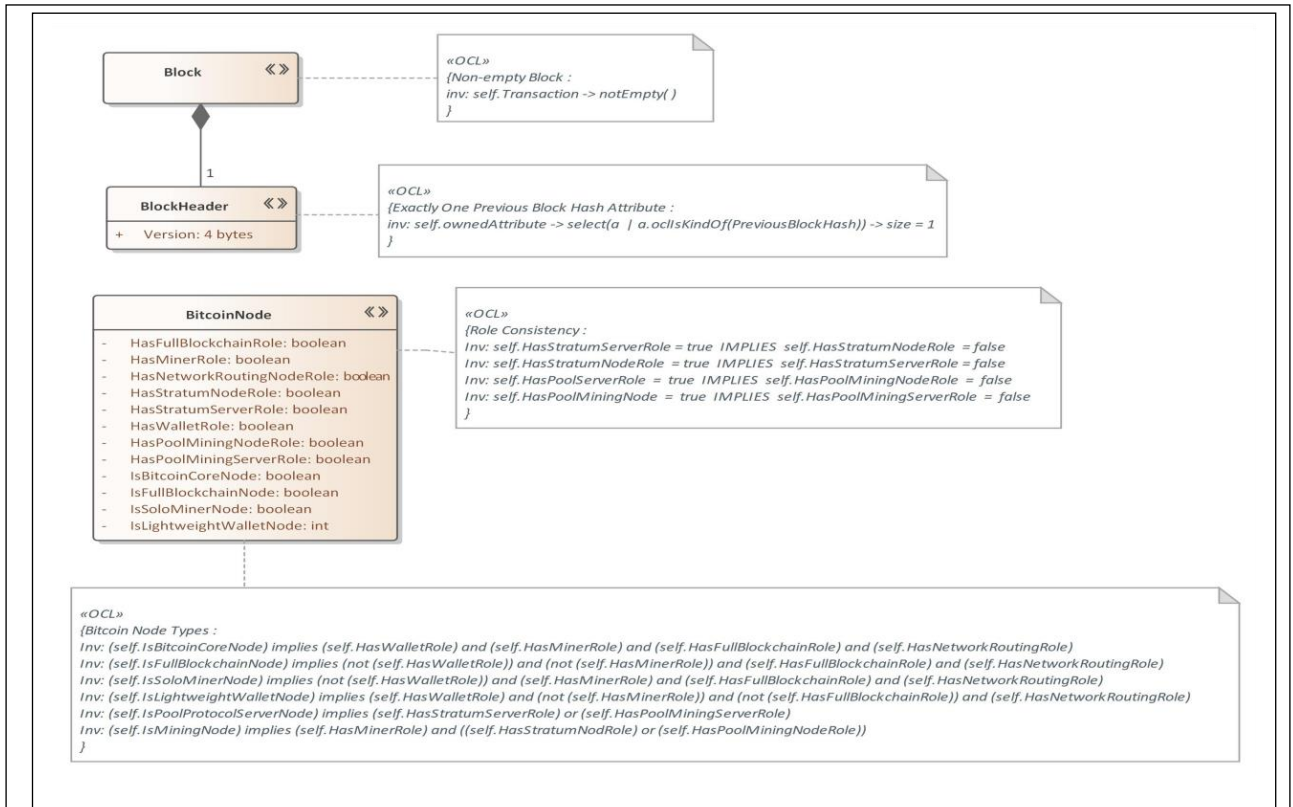

(a)

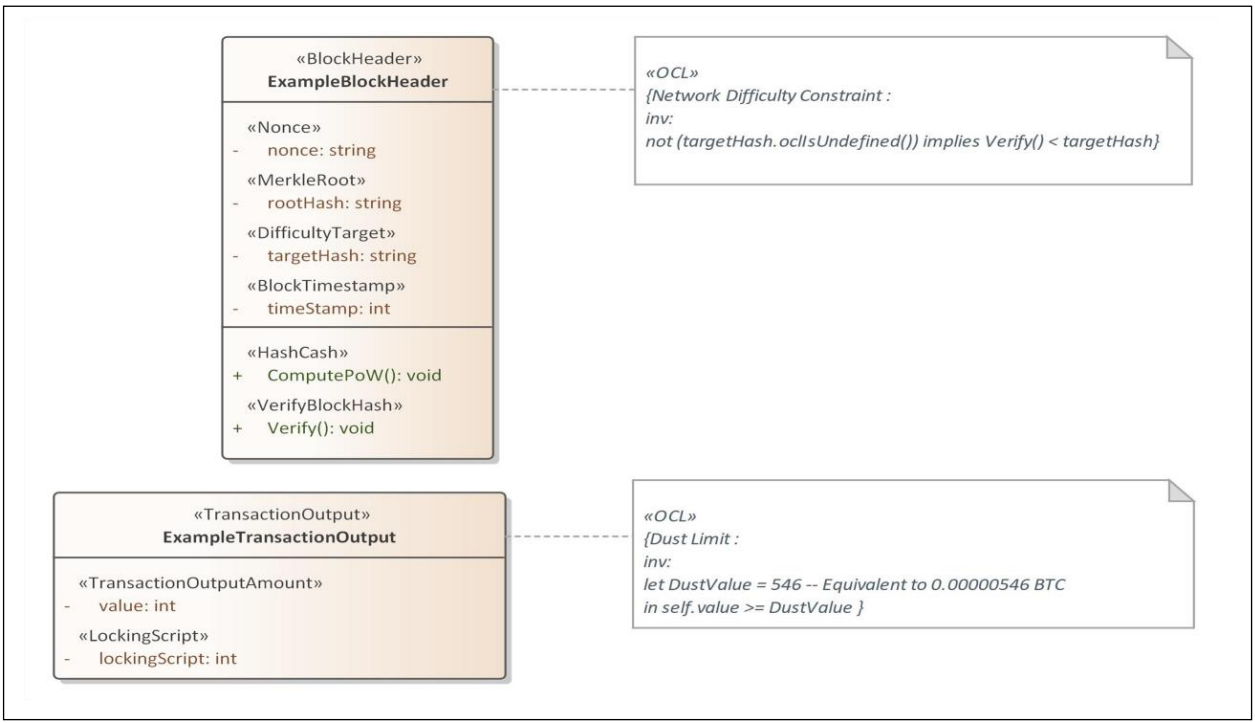

(b)

**Fig. 4.** (a) Example metamodel level OCL constraints; (b) Example application level constraints

As Fig. 6 shows, a tool-level constraint is defined to constrain classes stereotyped as *TransactionOutput* and *AbstractTransactionInput* as the allowed source and destination of the *Spend* connector, respectively. Hence, the *Spend* relationship can be drawn solely from classes stereotyped as *TransactionOutput* to classes stereotyped as *AbstractTransactionInput*, *TransactionInput,* or *CoinbaseTransactionInput*. In the same way, a constraint is defined to constrain classes stereotyped as *UnlockingScript* and *LockingScript* as the allowed source and destination of the *Unlock* relationship, respectively. In the same manner, the *Unlock* relationship can be drawn solely from classes stereotyped as *UnlockingScript* to classes stereotyped as *LockingScript*. As Fig. 5 depicts, the *Spend* connector extends the *UML Association* and shows that transaction inputs will provide proof of ownership (i.e. unlocking script) of the referenced UTXOs. The *Unlock* relationship extends the *UML Dependency* and is self-explanatory. In the same way, the remaining connectors (i.e. *PBKDF2KeyStretching*, *HMAC-SHA512,* and *RIPEMD160HashOfSHA256Hash* connectors) are defined in Fig. 5 and constrained in Fig. 6 and are used for bitcoin key management capabilities that may be used by a wallet or an exchange to access bitcoin blockchain and/or generate new transactions (Here wallet term is used to refer to Hierarchical Deterministic (HD) wallets as the most common type of bitcoin wallets). The *PBKDF2KeyStretching* connector can be used to model generating *seeds* from *mnemonic code word*s that are presented by *Seed* and *MnemonicCodeWord* stereotypes. Bitcoin uses 12-to-24-word mnemonic phrases. A wallet may use a mnemonic code word to generate a seed from which user private keys will be generated. The *HMAC-SHA512* connector models HMAC function which uses SHA512 hash function and its usage in bitcoin wallets is twofold: to derive a private key from a seed and a child private/public key from a parent private/public key. Finally, *RIPEMD160HashOfSHA256Hash* connector model double hashing used by bitcoin to generate *public addresses* from *public keys*. To this end, first, the *SHA256* hash of the public key is computed, and the result is then hashed using the *Ripmed160* hash function. This double hashing is modeled by *RIPEMD160HashOfSHA256Hash* connector and may be used to compute a public address *PublicAddress* = Ripmed160(SHA256(*PublicKey*)).

*MnemonicCodeWord, Seed, PrivateKey, PublicKey,* and *PublicAddress* stereotypes along with the last three described connectors (i.e. *PBKDF2KeyStretching*, *HMAC-SHA512,* and *RIPEMD160HashOfSHA256Hash* connectors), provide key management capabilities from which applications such as wallets and exchanges may benefit.

Bitcoin network nodes play different roles, including *wallet*, *miner*, *full blockchain*, *network routing*, *stratum node*, *stratum server*, and *pool server* [15]. Our profile includes a *BitcoinNode* stereotype, 12 tagged values, and a *CommunicationProtocol* enumeration which may be used in *BitML Network* diagrams to model application deployment and communication over the bitcoin network. These roles are specified as tagged values for the *BitcoinNode* stereotype and present functionalities. However, not all combinations of these roles are allowed. For example, no nodes are allowed to play *stratum node* and *stratum server* roles at the same time as defined by OCL constraint presented in Fig 4 (a). Furthermore, common combinations of the mentioned roles are defined as node types. For example, a node which plays *wallet*, *miner*, *full blockchain*, and *network routing* roles is called a *bitcoin core* node (or a *reference client* node). As another example, a node which plays *wallet* and *network routing* roles is called a *lightweight wallet* node. OCL corresponds to these constraints is presented in Fig. 4 (a). In addition to *BitcoinNode* constraints, two other meta-model level constraints are provided in Fig. 4 (a). The first avoids empty block creation, and the second enforces the use of exactly one previous block hash attribute. Finally, the application level constraint may be used; two examples of these constraints are provided in Fig. 4 (b). The first enforces the rule that the block hash of each valid block must be smaller than the network's difficulty target value, and the second enforces the dust limit of 546 Satoshis.

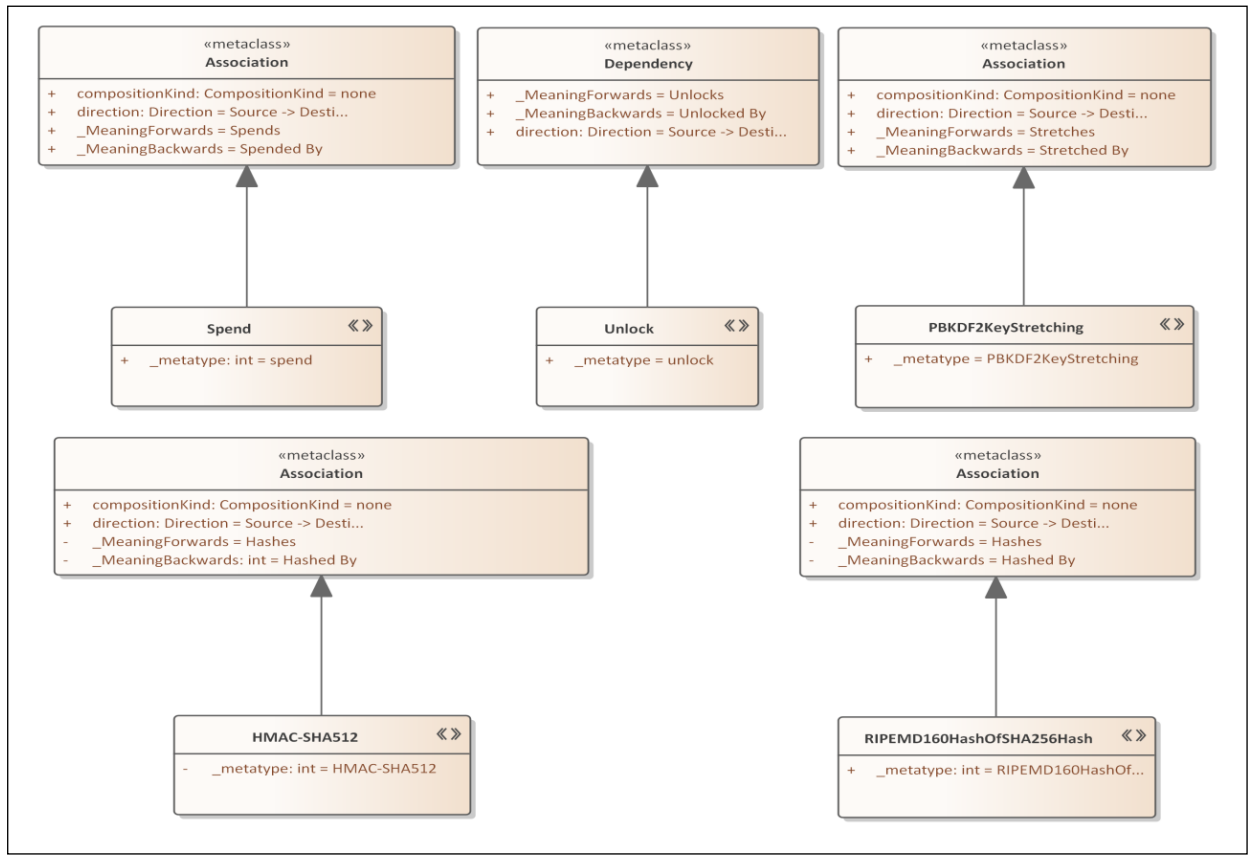

**Fig. 5.** Connector definition meta-model

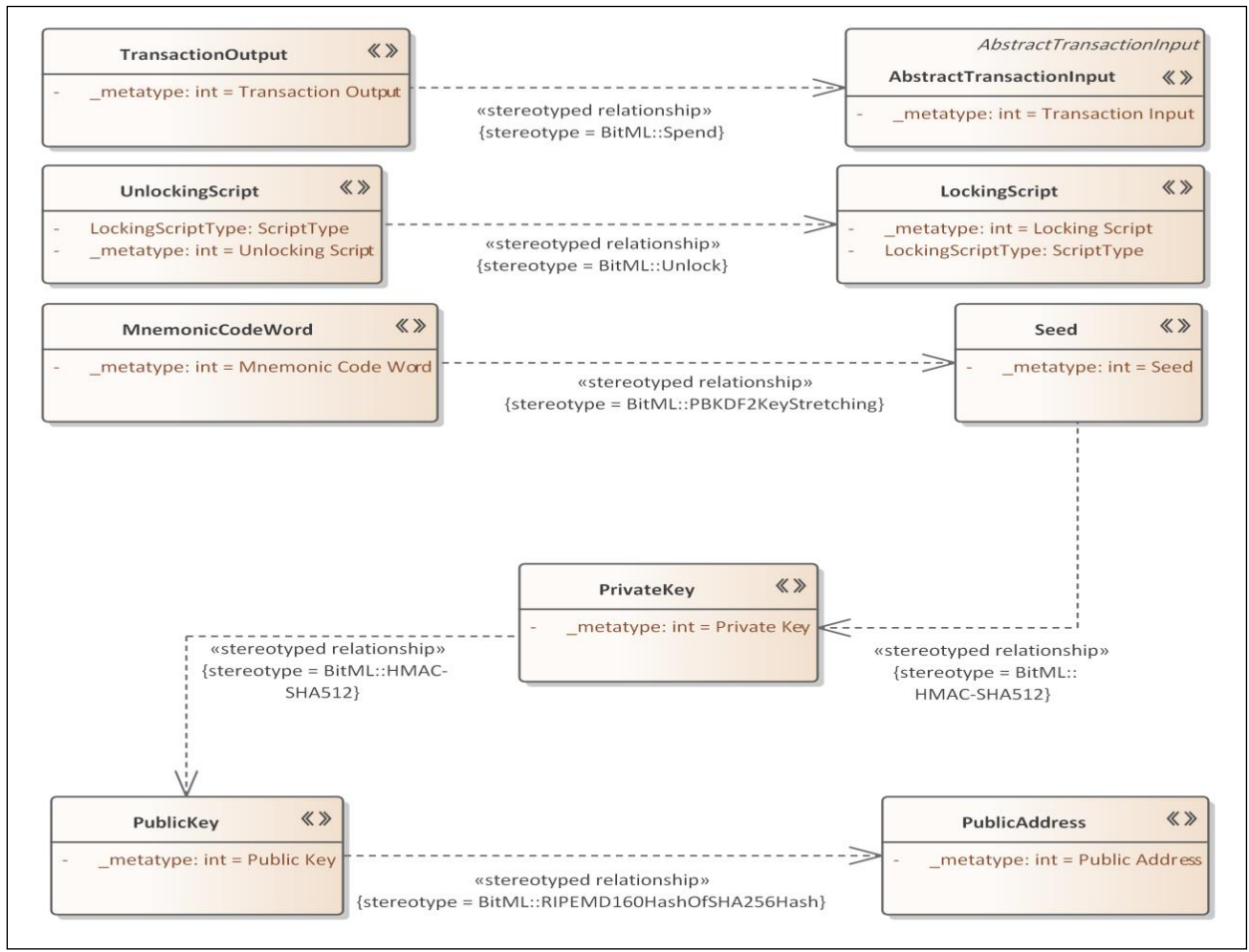

**Fig. 6.** Constraining connector usage

## 3.2. Implementation

The proposed profile has been implemented in the Sparx EA modeling tool. Fig. 7 shows a snippet example of XML generated by *Sparx MDG technology* for the defined UML profile. It shows that the *Unlock* relationship and the *UnlockingScript* stereotype extend the *UML Dependency* and *Class* meta-class, respectively. Furthermore, it demonstrates the applicability of the *Unlock* relationship on *UnlockingScript* stereotype and the definition of the *LockingScriptType* tagged value and its possible values.

As mentioned and depicted in Fig. 8, the *Transaction Processing* diagram and *Network* diagram extends the *Logical diagram* (i.e. *UML class diagram*) and *Deployment diagram*, respectively. Fig. 9 and Fig. 10 show definition and presentation of toolbox defined to provide access to BitML class stereotypes, BitML enumerations, and BitML connectors. Here, class stereotype means a stereotype that extends *Class* meta-class. Each element (i.e. class stereotype, enumeration, or connectors) is bound to the BitML toolbox by defining a tagged value for the extended toolbox. Depicted in Fig. 8, the BitML toolbox is assigned to the *Transaction Processing* diagram and *Network* diagram through the *toolbox* attribute of *Diagram_Logical* and *Diagram_Deployment* meta-classes, respectively. The appearance of the defined toolbox shown in Fig. 10 aims to ease access to defined stereotypes, enumerations, and connectors. Interchangeably, people may use the provided toolbox or directly stereotype the base elements (i.e. class or node). It is worth mentioning that we did not bind attribute/operation stereotypes to the toolbox, because they are not used directly; instead, they are used through their associated class stereotypes (i.e. they are attributes/operations of classes). Here, attribute and operation stereotypes mean stereotypes that extend *Attribute* and *Operation* meta-classes, respectively.

```xml
<Stereotype name="Unlock" metatype="unlock" notes="" cx="0" cy="0" bgcolor="-1" fontcolor="-1" bordercolor="-1"
        borderwidth="-1" hideicon="0">
    <AppliesTo>
        <Apply type="Dependency">
            <Property name="_MeaningForwards" value="Unlocks"/>
            <Property name="_MeaningBackwards" value="Unlocked By"/>
            <Property name="direction" value="Source -> Destination"/>
        </Apply>
    </AppliesTo>
</Stereotype>
<Stereotype name="UnlockingScript" alias="Unlocking Script" metatype="Unlocking Script" notes="" cx="0" cy="0"
        bgcolor="-1" fontcolor="-1" bordercolor="-1" borderwidth="1" hideicon="0">
    <stereotypedrelationships>
        <stereotypedrelationship stereotype="BitML::Unlock" constraint="BitML::LockingScript"/>
    </stereotypedrelationships>
    <AppliesTo>
        <Apply type="Class">
            <Property name="isActive" value=""/>
            <Property name="_HideUmlLinks" value="false"/>
        </Apply>
    </AppliesTo>
    <TaggedValues>
        <Tag name="LockingScriptType" type="enumeration" description="" unit="" values="P2PK,P2PKH,P2SH,P2WPKH" default=""/>
    </TaggedValues>
</Stereotype>
```

**Fig. 7.** An example of generated XML snippets

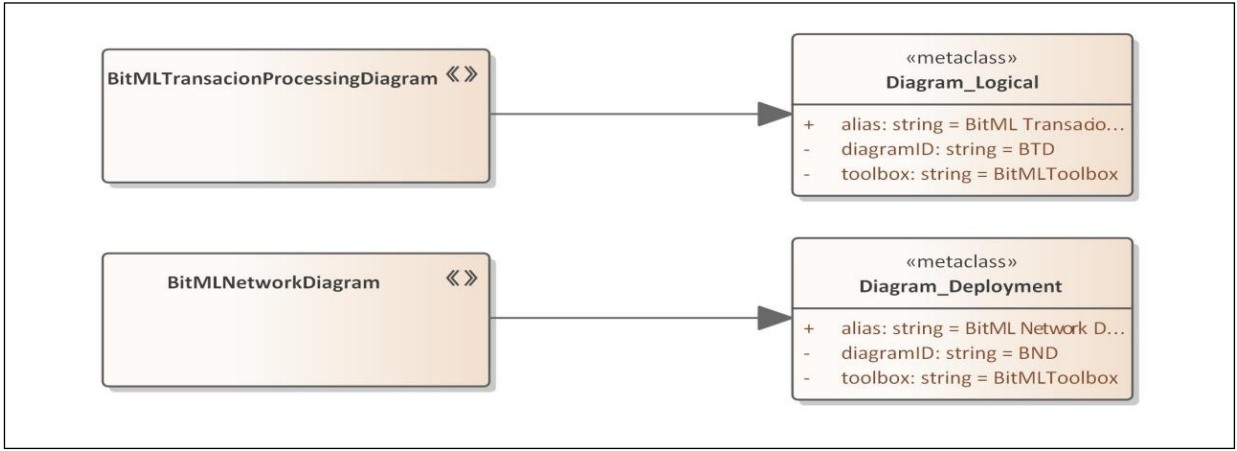

**Fig. 8.** Diagram definition meta-model

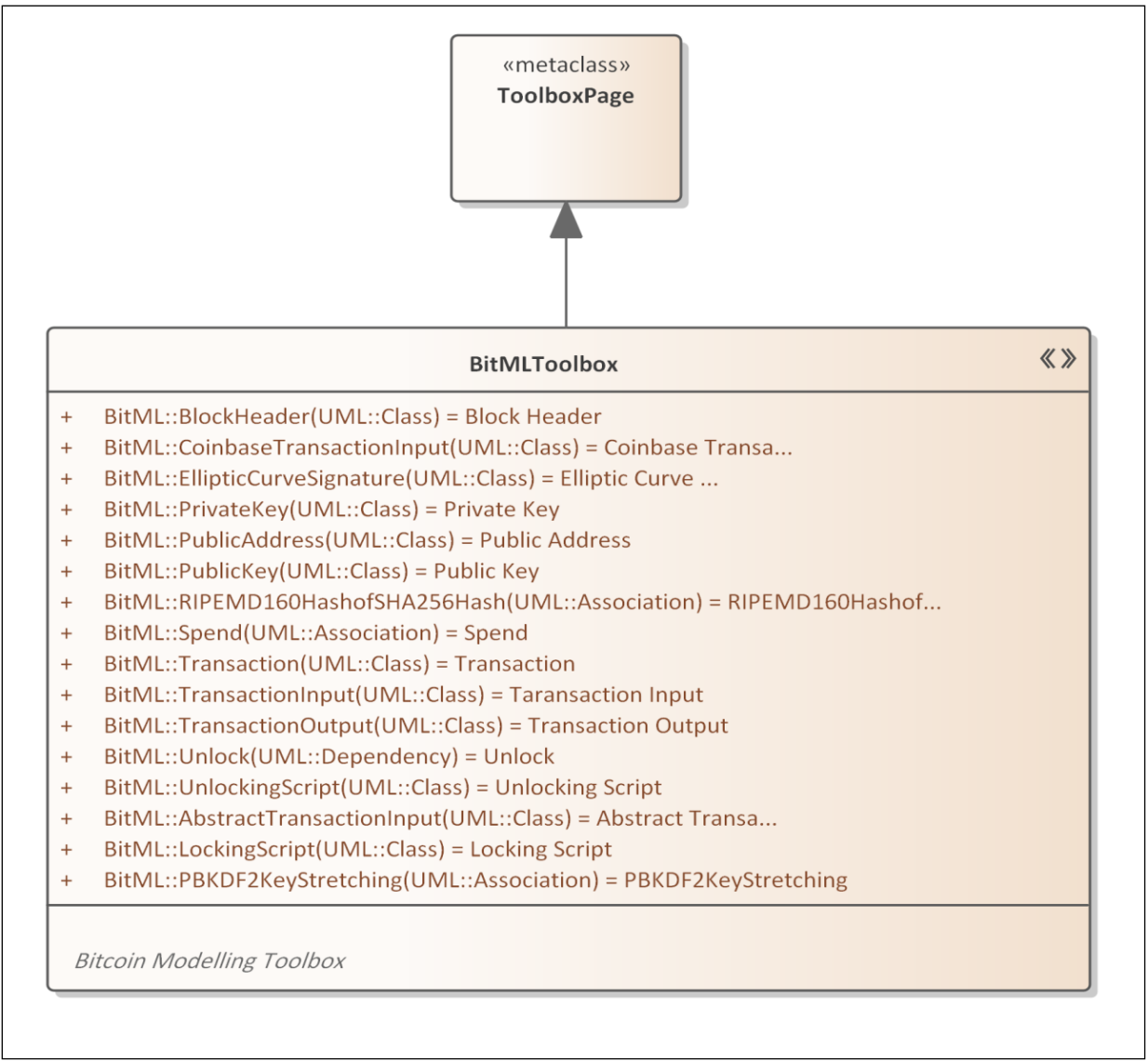

**Fig. 9.** Toolbox definition meta-model

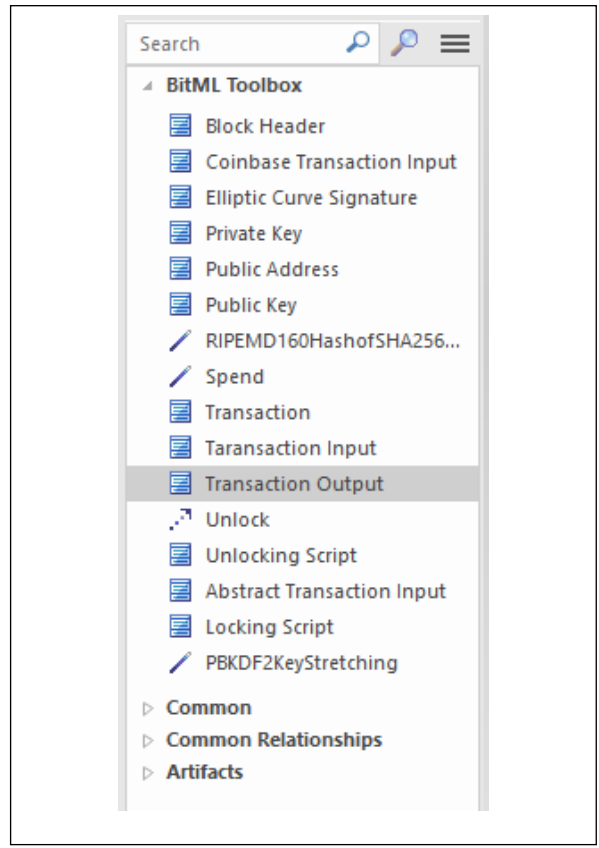

**Fig. 10.** Appearance of defined toolbox in Sparx EA

## 4. Case Study

As a secure bitcoin payment require at least 6 block confirmations, bitcoin is not used for payments that rely on quick transaction confirmation. Bitcoin fast payment algorithms [31, 32] aim to accelerate the payment process while resisting double spending. They provide a large number of off-chain transactions recorded by a small number of on-chain transactions. In an ongoing research project, we designed a fast payment application, implemented it using the NBitcoin library[2] and tested it on the Bitcoin Testnet[3]. Presented UML profile has been applied to model this application. This section presents one of these models along with defined constraints to illustrate how BitML can be used to model applications belonging to bitcoin application domain.

To this end, we used a *bond transaction* to create a specific UTXO which can be used for off-chain payments and resists double spending. A *refund transaction* which spends that UTXO in total after a specific period of time was defined by a transaction-level lock time. Moreover, fast payment application may generate some *settlement transactions,* which move funds from the fast payment account to merchants' accounts. A *settlement transaction* can produce one or two outputs: one, paying some bitcoins to your desired recipient, and optionally another output paying the remaining bitcoins in change back to your wallet. Each *settlement transaction* spends UTXO of either a *bound transaction* or a *change-back settlement transaction*.

---

[2] https://github.com/MetacoSA/NBitcoin
[3] https://en.bitcoin.it/wiki/Testnet

We modeled this application using the UML profile proposed in this paper. Our algorithms were modeled using *BitML Transaction Processing* diagrams and corresponding UML sequence diagrams. On-chain transactions of our fast payment application are modeled in Fig. 11. In addition to profile-defined constraints, Fig. 11 presents two application-level constraints. The first constraint, entitled "Lock Time Validation," constrains *bond transaction* and *refund transaction* to use zero and non-zero transaction level lock times, respectively. The second constraint, entitled "Subsequent Transaction Constraint," ensures that each *settlement transaction* spends UTXO of either a *bound transaction* or a *change-back settlement transaction* and has no transaction level lock time (zero) or a smaller transaction level lock time than the referenced one. This constraint ensures that the referenced UTXO cannot be spent beforehand. This model, which is an example of a *BitML Transaction Processing* diagram, along with other models was used in the analysis and design of the three mentioned algorithms, thus helping us to develop and communicate these three bitcoin fast payment algorithms without ambiguity.

## 5. Conclusion and Future Works

Blockchain has spread to nearly every industry, and different types of blockchain exists (i.e. public, private, consortium, hybrid, fourth generation, and fifth generation). Bitcoin is popular and highly valued and powered by a public blockchain, so people are encouraged to use more applications over the bitcoin blockchain. As bitcoin emerged as a convergence of distributed computing and mathematical cryptography, however, developing applications that run over the bitcoin blockchain became a challenging task, and the development of blockchain-specific software engineering best practices is sorely needed. To this end, the current paper proposed a UML profile for the bitcoin blockchain which includes two types of diagrams (i.e. *Transaction Processing* and *Network*), 42 stereotypes (16 for classes, 14 for attributes, eight for operations, and five for connectors), 23 tagged values, and six datatypes, and it defines a set of OCL constraints which allow automatic evaluation of conformance. From the provided background (i.e. bitcoin blockchain and UML profiling in Section 2), it can be concluded that instead of developing a modeling language from scratch, a UML profile such as our proposed work can benefit from both existing UML diagramming elements (i.e. diagrams, connectors, and so on) and profile specific diagramming elements to cope with and best model the intricate nature of application development over the bitcoin blockchain.

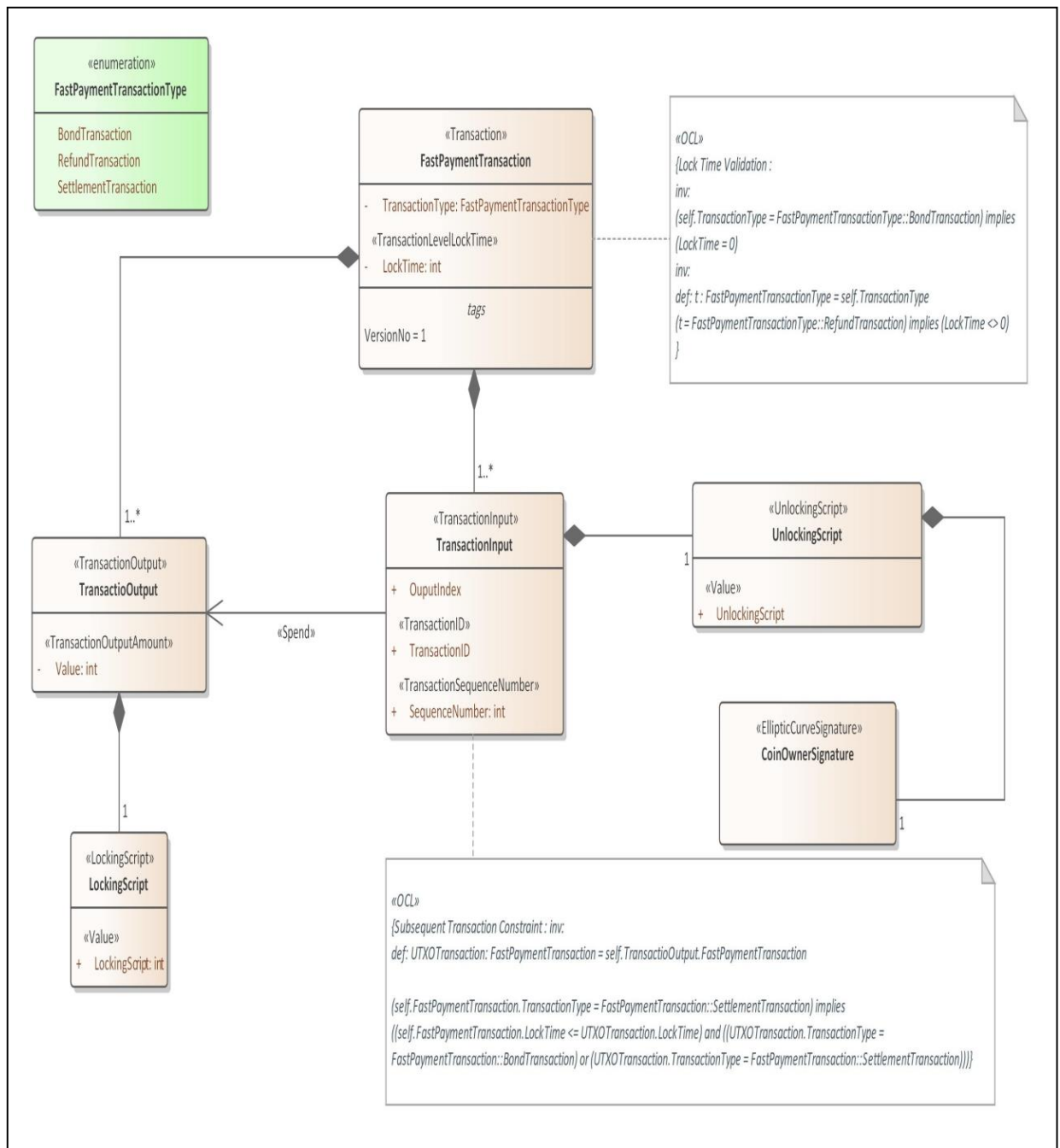

**Fig. 11.** An Example BitML Transaction Processing diagram for On-chain transactions of a bitcoin fast payment application

The proposed profile was implemented in *Sparx EA*. In addition, to illustrate how BitML can be used to model applications belonging to bitcoin application domain, a practical example developed by the proposed UML profile was presented as a case study.

Further research can facilitate side-chain development, and we aim to extend this work to support side-chain modeling. The authors expect that this extension, along with the other mentioned benefits, will increase the number of rational incentives for the wider adoption of the proposed UML profile and will encourage researchers, practitioners, and decision-makers to conduct experiments.